\documentclass[prb,showpacs,twocolumn,superscriptaddress]{revtex4}
\usepackage{graphicx}
\begin{document}

\title{Unconventional Superconducting Symmetry in a Checkerboard Antiferromagnet}

\author{Huai-Xiang Huang}
\affiliation{Department of Physics, Zhejiang University, Hangzhou,
310027, China} \affiliation{Department of Physics, Shanghai
University, Shanghai, 200444, China}
\author{You-Quan Li}
\affiliation{Department of Physics, Zhejiang University, Hangzhou,
310027, China}
\author{Jing-Yu Gan}
\affiliation{Center for Advanced Study, Tsinghua University,
Beijing, 100084, China}
\author{Yan Chen}
\affiliation{Department of Physics and Center of Theoretical and
Computational Physics, The University of Hong Kong, Pokfulam Road,
Hong Kong, China}
\author{Fu-Chun Zhang}
\affiliation{Department of Physics and Center of Theoretical and
Computational Physics,  The University of Hong Kong, Pokfulam Road,
Hong Kong, China} \affiliation{Department of Physics, Zhejiang
University, Hangzhou, 310027, China}

\date{\today}

\begin{abstract}
We use a renormalized mean field theory to study the Gutzwiller
projected BCS states of the extended Hubbard model in the large $U$
limit, or the $t$-$t'$-$J$-$J'$ model on a two-dimensional
checkerboard lattice. At small $t'/t$, the frustration due to the
diagonal terms of $t'$ and $J'$ does not alter the
$d_{x^2-y^2}$-wave pairing symmetry, and the negative (positive)
$t'/t$ enhances (suppresses) the pairing order parameter. At large
$t'/t$, the ground state has a $s-s$ wave symmetry. At the
intermediate $t'/t$, the ground state is $d+id$ or $d+is$-wave with
time reversal symmetry broken.
\end{abstract}


\maketitle

\section{Introduction}

Geometrically frustrated systems with strong correlation have
attracted much attention due to the highly nontrivial interplay
between frustration and correlation~\cite{book,Aoki}. In such
systems the pairwise interaction does not coincide with the geometry
of the lattice, which may lead to exotic ground states. In
particular, frustration in quantum magnets may cause certain types
of magnetically disordered quantum phases, including the resonating
valence bond (RVB) spin liquid state~\cite{Anderson87} and the
valence bond crystal state~\cite{VBC}. The quantum spin liquid state
could become unconventional superconducting state when the charge
carriers are introduced. There have been experimental evidences for
the unconventional superconductors in these systems. Examples are
the triangular layer cobaltates compound Na$_x$CoO$_2$~\cite{NaCoO},
layered organic conductor
$\kappa$-(ET)$_2$Cu$_2$(CN)$_3$~\cite{Org1}, the Kagome compound
SrCr$_8$Ga$_4$O$_{19}$~\cite{Kagome}, and 3-dimensional (3D) beta
type transition-metal pyrochlore material
KOs$_2$O$_6$~\cite{pyro1,pyro2}.
\begin{figure}
\includegraphics[width=4.8cm]{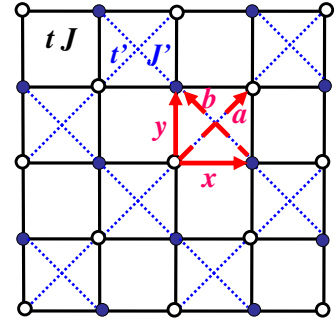}
\caption{\label{Fig1} (Color online). Schematic structure of 2D
checkerboard lattice. Solid (open) circles represent sublattice A
(B). The hopping integral and spin-spin coupling are $t$ and $J$ for
the nearest neighbor links (solid lines), and $t'$ and $J'$ for the
diagonal or next nearest neighbor links (dashed lines),
respectively.}
\end{figure}

To describe the interplay between frustration and correlation, we
consider a $t$-$t'$-$J$-$J'$ model on a 2D checkerboard lattice, and
analyze the possible superconducting pairing symmetry of the model.
The checkerboard lattice is a frustrated one, and may be considered
as a 2D projection of a 3D corner-sharing lattice of pyrochlore. A
schematic checkerboard lattice is illustrated in Fig. 1. The
Hamiltonian reads,
\begin{eqnarray}
H = &-&\sum_{\langle ij \rangle \sigma} t_{ij} P_D (c_{i\sigma
}^{\dagger }c_{j\sigma }+ h.c.)P_D+ \sum_{\langle ij \rangle}J_{ij}
\vec{S}_{i}\cdot \vec{S}_{j} \nonumber \\
 & - & \mu\sum_{i \sigma} c_{i\sigma
}^{\dagger }c_{i\sigma }
\end{eqnarray}
where $c_{i\sigma }^{\dagger }$ is an electron creation operator
with spin $\sigma $ at site $i$, $\vec{S}_{i}$ is a spin operator
for electron, $\mu$ is the chemical potential, and $\langle ij
\rangle$ denotes a neighboring pair on the lattice. $P_D$ is a
Gutzwiller projection operator to impose no double electron
occupation at any site on the lattice. $t_{ij}$ and $J_{ij}$ stand
for the hopping integrals and antiferromagnetic exchange couplings
respectively, and $t_{ij}=t$, $J_{ij}=J$ for the nearest neighbor
(n.n) links and $t_{ij}=t'$, $J_{ij}=J'$ for the diagonal or the
next n.n. links as shown in Fig. 1. For convenience, we use $x, y$
to represent the n.n. links while $a,b$ to describe the two diagonal
links. The Hamiltonian may be viewed as a strong coupling limit of a
Hubbard model on the lattice with n.n. and next n.n. hopping
integrals $t$ and $t'$ respectively and a on-site Coulomb repulsive
interaction $U$. Hereafter we use $t$ as an energy unit and set
$J/t=1/3$. We choose $J'/J=(t'/t)^{2}$, consistent with the
superexchange relation of $J= 4t^2/U$ in the large $U$ limit of the
Hubbard model.

The model has certain limiting cases. At $t'/t \rightarrow 0$, the
model is reduced to the $t$-$J$ model on a square lattice. At $t'/t
\rightarrow \infty$, the model becomes a collection of independent
1D $t'$-$J'$ chains. At $t'/t=1$, it is an isotropic checkerboard
lattice with a highly geometrically frustrated structure.

Previous theoretical investigations mainly focused on the
half-filled case without charge carriers. A variety of techniques
have been employed to study the quantum antiferromagetism on such a
lattice~\cite{Fuji,paradigm,olegt,hermele,Balents05}. Various
quantum paramagnetic ground states may appear, which include some
translational symmetry breaking states and a quantum spin liquid
state. The introduction of doping with mobile charge carriers in a
frustrated quantum antiferromagnet may result in the appearance of
unconventional
superconductivity~\cite{Ogata03,d+id,Kim04,palee,Gan06}. Generally
speaking, geometric frustration may play a key role in the mechanism
of unconventional superconductivity. Recently exact diagonalization
approaches have been employed to study the superconducting
fluctuations in this system with $t'$=$t$ and
$J'$=$J$~\cite{Poil1,Poil2,Poil3}. They found evidence of
enhancement in pairing amplitude at arbitrarily small $J/t$ for a
specific sign of the hopping amplitude.

In this paper, we apply the plain vanilla version of the RVB
theory~\cite{Zhang88,vannila} to study the ground state of the
$t$-$t'$-$J$-$J'$ model on a 2D checkerboard lattice. The
competition among various superconducting states will be examined.
Since our primary interest is on the possible pairing symmetry of
the superconducting state for the doped system, we will not consider
the possible long-range magnetic ordering in the present paper. Our
main results can be summarized below. The $d_{x^2-y^2}$-wave pairing
found for the $t$-$J$ model extends to a large region in parameter
space of $t'/t$ and doping concentration $\delta$, and the negative
$t'/t$ enhances the pairing while the positive $t'/t$ suppresses the
pairing. At small doping and for $|t'/t| \sim 1$, there is a small
region where the pairing symmetry is $d+id$ or $d+is$. At $|t'/t|
>1$, there are regions where the pairing symmetry belongs to a
$s-s$ wave.

The rest of the paper is organized as follows. In Sec. II, we apply
the renormalized mean field theory to study the RVB state. In Sec.
III, we present our numerical results on the possible
superconducting ground states. In particular, four distinct
superconducting phases show up in the phase diagram as a function of
$t'/t$ and the doping $\delta$.  Sec. IV is a summary. The
diagonalization of the mean-field Hamiltonian and the explicit form
of the self-consistent equations are presented in the Appendix.

\section{Formalism}
We use a Gutzwiller projected BCS state\cite{Anderson87} as a trial
wavefunction to study the ground state and the corresponding pairing
symmetry of the Hamiltonian (1). The trial wavefucntion is of the
form,
\begin{eqnarray}
|\Psi _{GS} \rangle =\prod\limits_{i}
(1- n_{i\uparrow}n_{i\downarrow})|\Psi _{BCS}\rangle
\end{eqnarray}
where $n_{i,\sigma}=c^{\dag}_{i,\sigma}c_{i,\sigma}$, and the
projection operator $\prod\limits_{i}(1- n_{i\uparrow
}n_{i\downarrow })$ removes the doubly occupied electron states on
every lattice site $i$. We use a renormalized mean field theory
~\cite{Zhang88,vannila} to calculate the energy of the Hamiltonian
(1). In the renormalized mean field theory, we adopt the Gutzwiller
approximation to replace the projection by a set of renormalized
factors, which are determined by statistical
countings~\cite{gutzwiller,vallhardt}. We have
\begin{eqnarray}
\langle c_{i\sigma }^{\dagger }c_{j\sigma }\rangle  = g_{t}\langle
c_{i\sigma }^{\dagger }c_{j\sigma }\rangle _{0} , ~~~
\langle \vec{S}_{i}\cdot \vec{S}_{j}\rangle  =g_{s}\langle \vec{S}%
_{i}\cdot \vec{S}_{j}\rangle _{0}
\end{eqnarray}
where $g_{t}$ and $g_{s}$ are the Gutzwiller renormalized
factors and are given by~\cite{Zhang88} $g_{t}=2\delta /(1+\delta)$
and $g_{s}=4/(1+\delta)^2$ where $\delta$ denotes the doping
density. Therefore, the variational calculations of $H$ of Eq. (1) in the projected
BCS state is reduced to the variational calculations of an effective Hamiltonian $H_{eff}$
given below in the
unprojected BCS states $|\Psi _{BCS} \rangle$ for $H_{eff}$ .
\begin{eqnarray}
H_{eff} & = & \sum_{\langle ij \rangle \sigma} -g_t t_{ij}
(c_{i\sigma }^{\dagger }c_{j\sigma } + h.c) +\sum_{( ij )} g_s
J_{ij}
\vec{S}_{i}\cdot \vec{S}_{j} \nonumber \\
 & - & \mu\sum_{i \sigma} n_{i \sigma}.
\end{eqnarray}

To proceed further, we introduce particle-particle and particle-hole
mean fields, ($\tau= \pm \hat{x}, \pm \hat{y},\pm \hat{a},\pm
\hat{b}$, see Fig. 1)
\begin{eqnarray}
{\Delta}_\tau &=& \langle
c^{\dag}_{i\uparrow}c^{\dag}_{i+\tau\downarrow}-
c^{\dag}_{i\downarrow}c^{\dag}_{i+\tau\uparrow}\rangle_0 \nonumber
\\
\xi_\tau &=& \sum_{\sigma}\langle
c^{\dag}_{i\sigma}c_{i+\tau,\sigma}\rangle_{0}.
\end{eqnarray}
Here we focus on the translational invariant state with the spin
singlet and even parity superconducting pairing symmetry, where
$\Delta_{i,i+\tau} = \Delta_{i+\tau, i} =\Delta_{\tau}$, and
$\chi_{i,i+\tau}=\chi_{\tau}$. The superconducting order parameter
is a $2\times2$ matrix representing the two sublattices,
\begin{eqnarray}\label{matrix}
 \Delta_{SC}(\vec{k})=\left(%
\begin{array}{cccc}
\langle c_{\vec{k} \uparrow A}^{\dagger} c_{\vec{-k} \downarrow
A}^{\dagger} \rangle & \langle c_{\vec{k} \uparrow A}^{\dagger}
c_{\vec{-k} \downarrow B}^{\dagger} \rangle
\\
\langle c_{\vec{k} \uparrow B}^{\dagger} c_{\vec{-k} \downarrow
A}^{\dagger} \rangle & \langle c_{\vec{k} \uparrow B}^{\dagger}
c_{\vec{-k} \downarrow B}^{\dagger} \rangle\\

\end{array}%
\right)
\end{eqnarray}
Within the Gutzwiller approximation, it is related to
$\Delta_{\tau}$ by
\begin{eqnarray}\label{matrix}
 \Delta_{SC}(\vec{k})= g_t \left(%
\begin{array}{cccc}
\Delta_a cosk_a  & \Delta_x cosk_x + \Delta_y cosk_y
\\
\Delta_x cosk_x + \Delta_y cosk_y  & \Delta_b cosk_b\\
\end{array}%
\right)
\end{eqnarray}
In the limit of $\delta=0$, consistent with the fact that the ground
state is a Mott insulator at half filling. At small $\delta$, we
have $\Delta_{SC}(\vec{k}) \sim \delta$.

In terms of these mean fields, the effective Hamiltonian can
be expressed as
\begin{eqnarray}
H_{MF}&=&\sum_{ \langle ij \rangle \sigma}-\frac{3}{8} g_s J_{ij} [\xi
_{ij} c_{i\sigma }^{\dagger }c_{j\sigma } + \Delta _{ij} c_{i
\sigma}c_{j \bar{\sigma}} + h.c. ]\nonumber\\
&\,&-g_t t_{ij} (c_{i\sigma }^\dag  c_{j\sigma } + h.c) - \mu\sum_{i
\sigma} n_{i \sigma} + const.
\end{eqnarray}

There are eight independent complex mean-field parameters: $\xi_x,
\xi_y, \Delta_x, \Delta_y$ on the n.n. links and $\xi_a,
\xi_b,\Delta_a, \Delta_b$ on the next n.n. links. Here we assume all
the particle-hole mean fields $\xi$ to be real. We denote
$\theta_{x,y}$, $\theta_{a,b}$, and $\theta_{x,a}$ as the relative
phases of $(\Delta_x, \Delta_y)$, $(\Delta_a, \Delta_b)$, and
$(\Delta_x, \Delta_a)$, respectively. The energy per site can be
expressed in terms of the mean fields and is given by
\begin{eqnarray}\label{4plak}
E_{gs}&=&-2g_tt(\xi_x+ \xi_y)-g_tt'(\xi_a+\xi_b)\nonumber\\
&\,&-\frac{3}{8}g_sJ(\xi_x^2+\xi_y^2+|\Delta_x|^2+|\Delta_y|^2)\nonumber\\
&\,&-\frac{3}{16}g_sJ'(\xi_a^2+\xi_b^2+|\Delta_a|^2+|\Delta_b|^2).
\end{eqnarray}
The mean field parameters $\xi$ and $\Delta$ and the chemical
potential $\mu$ can be determined by solving the self-consistent Eq.
(5) together with an equation for the hole density. The mean field
state at zero temperature can be obtained by the diagonalization of
$H_{MF}$. We then determine the lowest energy state for each set of
parameters $t'/t$ and $\delta$. The detailed formalism of the
diagonalization of $H_{MF}$ and the self consistent equations can be
found in Appendix. These equations can be solved numerically, and
the results are given in the next section.
\begin{figure}[tb]
\includegraphics[width=8cm]{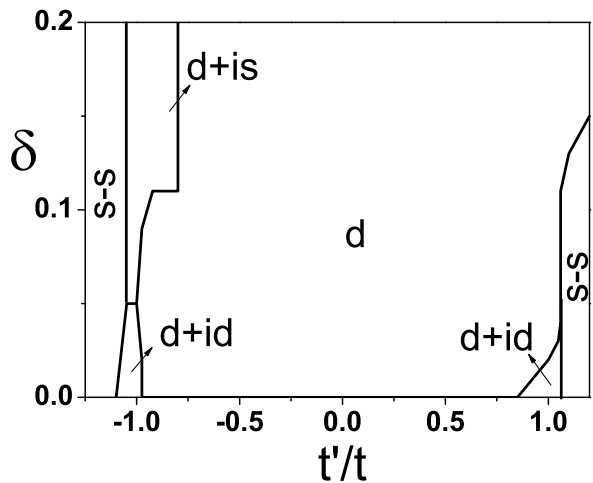}
\caption{\label{Fig2} Phase diagram of $t$-$t'$-$J$-$J'$ model on a
checkerboard lattice in parameter space of hole density $\delta$ and
$t'/t$ obtained by the renormalized mean field theory. The phases
are indicated by their superconducting pairing symmetry defined in
Table I, and $J'/J =(t'/t)^2$.}
\end{figure}

\section{Numerical Results}

\begin{table}
\caption{The pairing symmetries (shown in Fig. 2 and used in the
text) of the ground states of $t$-$t'$-$J$-$J'$ model on a
checkerboard lattice and their corresponding mean fields
$\Delta_{\tau}$ of Eq. (5). $\theta_{\tau,\tau'}$ are the relative
phase between $\Delta_{\tau'}$ and $\Delta_{\tau}$}.
\begin{tabular}{|c|c|}\hline
Pairing symmetry & mean field parameters \\
   \cline{1-2}
 $d$  & $\Delta_x=-\Delta_y$ , $\Delta_a=\Delta_b=0$ \\
  \cline{1-2}
 $d+id$  & $\Delta_x=-\Delta_y$, $\Delta_a=-\Delta_b$, $\theta_{x,a} \approx \pi/2$ \\
  \cline{1-2}
 $s-s$  & $\Delta_x=\Delta_y$, $\Delta_a=\Delta_b$, $\theta_{x,a}=\pi$  \\
  \cline{1-2}
 $d+is$  & $\Delta_x=-\Delta_y$, $\Delta_a=\Delta_b$, $\theta_{x,a} \approx \pi/2$ \\
  \hline
\end{tabular}
\end{table}
In this section, we present the numerical results of the
self-consistent renormalized mean field theory for Hamiltonian Eq.
(1) on the checkerboard lattice. We will first discuss the phase
diagram, then provide detailed analyses of the mean field parameters
as functions of $\delta$ for several typical values of $t'/t$. The
phase diagram is shown in Fig. 2. The ground state at $\delta=0$ is
a Mott insulator. At finite doping, the ground state is a
superconducting state, with four different types of pairing
symmetry, as illustrated in Table 1. Here we classify the pairing
symmetry in terms of the relative phases between $\Delta_y$ and
$\Delta_x$, between $\Delta_b$ and $\Delta_a$, and between
$\Delta_a$ and $\Delta_x$. Such a classification is consistent with
the four-fold rotational symmetry in the Bravis lattice of the
checkerboard structure.

At the limit $t'$=$J'$=0, the model is reduced to the $t$-$J$ model,
and the ground state has a $d_{x^2-y^2}$ or $d$-wave
symmetry~\cite{Zhang88,kotliar} at finite doping. This pairing state
is robust against the next n.n. terms. As we can see from Fig. 2,
the $d$-wave pairing symmetry of the ground state extends to a large
region of both positive and negative values of $t'/t$. In such
state, $\Delta_a=\Delta_b=0$, but there is an additional self-energy
term arising from the next n.n. spin coupling. There are nodal
quasiparticles, whose position are determined by the crossing of the
lines $\cos{k_x}=\cos{k_y}$ and the Fermi surface, similar to those
obtained in the $t$-$J$ model. Note that the $d$-wave pairing has
been previously found to be stable against the weak frustrations as
studied by various authors \cite{Ogata03,d+id,Gan06} on anisotropic
triangular lattices. At large $|t'/t|$, the ground state has a $s-s$
wave pairing symmetry with $\Delta_x=\Delta_y$ and
$\Delta_a=\Delta_b$. In that state, the relative phase between
$\Delta_a$ and $\Delta_x$ is $\pi$. Between the above two regions,
there is a small parameter region around $|t'/t|=1$, where the
ground state has a $d+id$ pairing symmetry at small $\delta$, and a
$d+is$ phase at larger $\delta$ for negative $t'/t$. In both $d+id$
and $d+is$ states, the relative phases between $\Delta_a$ and
$\Delta_x$ are close to $\pi/2$, and are weakly dependent of
$\delta$, and the time reversal symmetry is spontaneously broken. We
note that Hamiltonian (1) is asymmetric with respect to positive and
negative values of $t'/t$, which is reflected in our phase diagram.
Similar to the case in the $t$-$J$ model, the particle-particle mean
field amplitude $\Delta$ disappears at large $\delta$, and the
ground state becomes a normal metal. The details of the phase
boundaries between the superconducting and normal metallic states
will be elaborated below. We also note that in the limit $|t'/t|
\rightarrow \infty$, the ground state becomes 1D-like.

\begin{figure}[tb]
\includegraphics[width=8cm]{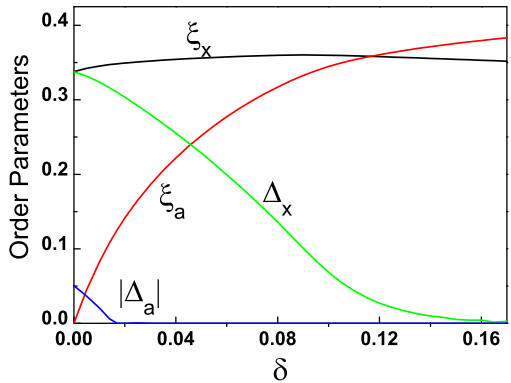}
\caption{\label{Fig3} (Color online). Amplitudes of mean fields
$\xi$ and $\Delta$ as functions of hole density $\delta$ for
parameters $t=t'=1, J=J'=1/3$. The ground state has $d+id$ pairing
(see Table 1) at $0<\delta < 0.02$, and $d_{x^-y^2}$-wave pairing at
$0.02 < \delta < 0.15$, and is a normal state at $\delta>0.15$.}
\end{figure}
In Fig. 3, we display the amplitudes of the mean field parameters as
functions of $\delta$ at the symmetric point $t'/t=1$. At very low
hole density, the ground state has $d+id$-wave pairing symmetry, and
$\Delta_a = i |\Delta_a|$, but $|\Delta_a|\ll \Delta_x$. As $\delta$
increases, the ground state becomes a $d$-wave, and
$\Delta_a=\Delta_b$ vanishes. In comparison with the mean field
amplitudes found in the $t$-$J$ model, $\xi_x$ is similar and
insensitive to $\delta$, but $\Delta_x$ decreases more rapidly in
the checkerboard model as $\delta$ increases. The latter may be
understood as the consequence of the non-zero value of $\xi_a$ in
the present case, which increases rapidly as $\delta$ increases.
Note that the $s-s$ wave state has a very close energy, although it
is slightly higher than either $d+id$- or $d$- wave states.
Recently, Poilblanc~\cite{Poil2} performed a finite cluster exact
diagonalization study of the $t$-$J$ model on a checkerboard
lattice. Some exotic states for positive $t$ are found to have
$d_{x^2-y^2}$-, $s-s$- symmetries, which appears to be consistent
with our results.

\begin{figure}[tb]
\includegraphics[width=8cm]{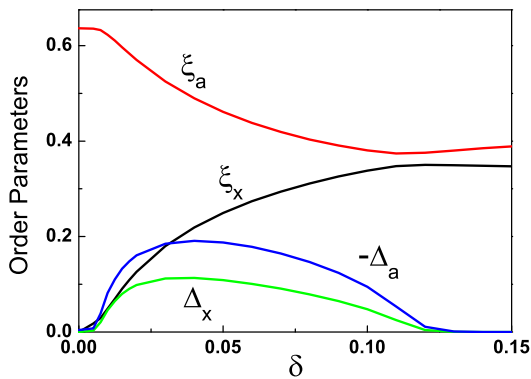}
\caption{\label{Fig4} (Color online). Amplitudes of the mean fields
$\xi$ and $\Delta$ as functions of $\delta$ for parameters $t=1,\,
t'/t=1.1, \, J=1/3,\, J'/J=1.21$. The ground state is a $s-s$ wave
at $0.02 < \delta < 0.12$, and a normal state otherwise.}
\end{figure}
The mean field amplitudes as functions of $\delta$ for the model at
$t'/t=1.1$ are depicted in Fig. 4. The ground state is a $s-s$ wave
at $0.02<\delta<0.12$. The pairing amplitudes disappear around
$\delta=0.12$, indicating the ground state to be a normal metallic
state at $\delta >0.12$. As we can see, the correlations along the
diagonal directions ($\hat a$ and $\hat b$) become more important
than those along $\hat x$ and $\hat y$ directions at $t'/t>1$. Note
that the results near the half filled need to be cautious. At the
half filling, the mean field ground state has only non-zero values
of $\xi_a$ and $\xi_b$, indicating that the state is a collection of
the decoupled chains along the directions of $\hat a$ and $\hat b$.
This may attribute to the poor Gutzwiller approximation on 1D
systems\cite{Zhang88}.

In Fig. 5 and Fig. 6, we show typical $\delta$ dependence of the
mean field amplitudes for parameters $t'/t <0$. In this case,
$\xi_a$ and $\xi_b$ have the opposite sign with $\xi_x$ or $\xi_y$.
At small value of $|t'/t|$, the ground state is a $d$-wave state,
the amplitudes of the mean fields are illustrated in Fig. 5 for
$t'/t=-0.5$. In that state, $\Delta_a=\Delta_b=0$, and $\xi_a=\xi_b$
are small and changes sign at $\delta > 0.2$. Interestingly,
$\Delta_x$ decreases much slower as $\delta$ increases, in
comparison with that in the $t$-$J$ model. This suggests that the
superconducting state may extend to a much larger hole
density.

\begin{figure}[tb]
\includegraphics[width=8cm]{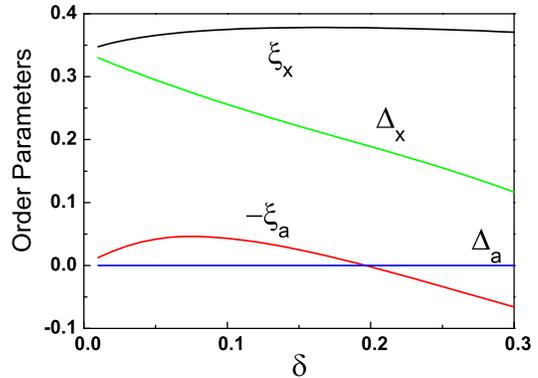}
\caption{\label{Fig5} (Color online). Amplitudes of mean fields
$\xi$ and $\Delta$ as functions $\delta$ in the case of $t=1,\,
J=1/3,\, t'/t=-0.5\, J'/J=0.25$. $\Delta_a=\Delta_b=0$, and the
ground state has a $d$-wave pairing symmetry(see Table 1).}
\end{figure}

In Fig. 6, we plot the mean field amplitudes for $t'/t=-1$. Away
from half-filling, the amplitudes of mean fields change
nonmonotonically and there exist three distinct pairing symmetries
with respect to different doping levels. As hole density increases,
the ground state evolves from the $d+id$-wave state
($\Delta_a=-\Delta_b=i|\Delta_a|$) to the $d$-wave state and to
$d+is$-wave state ($\Delta_a=\Delta_b=i|\Delta_a|$), as we can see
from the figure that the amplitude of $\Delta_a$ decreases to zero
around $\delta=0.02$ and arises again at $\delta=0.06$. In contrast
to the $d+id$-wave state, the amplitude of $\Delta_{a(b)}$ is
comparable to $\Delta_x$ in the $d+is$-wave state. In comparison
with the case of positive $t'/t$, we note that the suppression of
$\Delta_x$ becomes much slower as $\delta$ increases and the
superconductivity appears more favored for negative $t'/t$. This
result is in agreement with previous studies for a $t$-$J$ model on
a triangular lattice~\cite{Ogata03,d+id}. An intuitive physical
understanding of such effect can be given as follows: for positive
$t'/t$, the $t'$ and $t$ terms in kinetic energy match quite well,
so that $t'$ term may enhance the kinetic energy (make it lower),
hence suppress the pairing amplitude; for negative $t'/t$, the $t'$
term may introduce frustration in kinetic energy, hence enhances the
pairing amplitude. The situation here is similar to that in
cuprates, where the positive $t'/t$ case corresponds to the
electron-doped system while the negative $t'/t$ case corresponds to
the hole-doped system.  It is well known that the hole-doped system
has a higher transition temperature while the electron-doped system
has a lower one and a substantial region in doping with
antiferromagentic long range order which we have not considered in
the present paper for simplicity.

\begin{figure}[tb]
\includegraphics[width=8cm]{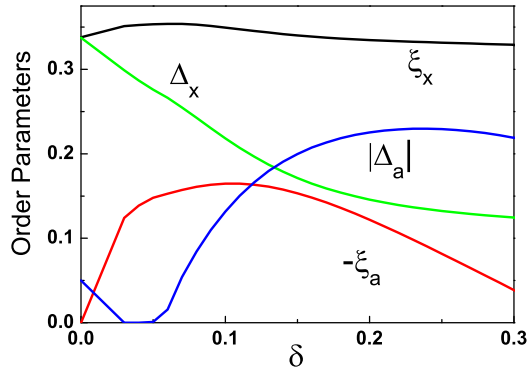}
\caption{\label{Fig6} (Color online). Amplitudes of mean fields
$\xi$ and $\Delta$ as functions $\delta$ for the parameters $t=1,\,
J=J'=1/3, \, t'=-1$. The ground state is $d+id$-wave at $0< \delta <
0.02$, $d$-wave at $0.02 <\delta < 0.06$, and $d+is$-wave at
$\delta> 0.06$.}
\end{figure}

\section{Summary}

We have applied the renormalized mean field theory to study the
Gutzwiller projected BCS ground state for the $t$-$t'$-$J$-$J'$
model on a highly frustrated checkerboard lattice. As charge
carriers are introduced, the half filled Mott insulator is evolved
to resonating valence bond superconducting states, with four types
of pairing symmetries depending on the hole density and the
parameter $t'/t$. We have found that the $d_{x^2-y^2}$ symmetry
state previously obtained for the $t$-$J$ model is most stable in a
large parameter region for small values of $|t'/t|$. In this region,
the $d$- wave pairing order parameter is suppressed for positive
$t'/t$ and enhanced for negative $t'/t$. At large value of $|t'/t|$,
the ground state is found to be so-called $s-s$ wave. Around
$|t'/t|=1$, we have found the time reversal symmetry broken states.
$d+id$- and $d+is$- symmetries states are the most stable. The
appearance of the exotic superconducting pairing symmetry in the
$t$-$t'$-$J$-$J'$ model on a checkerboard lattice may be viewed as
the interplay between geometrically frustration and strong electron
correlation. Our results may shed light on the understanding of the
novel pairing symmetry features of highly frustrated systems.

HXH thanks X. Wan for numerical assistance. YC would like to thank
Z.D. Wang and Q.H. Wang for helpful discussions. This research was
supported by the RGC grants (HKU-3/05C, HKU7057/04P and HKU7012/06P)
of Hong Kong SAR government, seed funding grant from the University
of Hong Kong, and NSF in China No.10225419 and No.10674117.

\section{Appendix}

In the appendix, we present detailed calculations of the mean field
theory. We first diagonalize the mean field Hamiltonian (6) on the
checkerboard lattice. We carry out the Fourier transformation of
electron operator in the two sublattices $A$ and $B$,

\begin{eqnarray}
c_{i\sigma }= \sum_{\vec{k}} \frac{1}{\sqrt{N}}a_{\vec{k}\sigma
}\exp(i\vec{k} \cdot \vec{r}_i),
\qquad i \in A \\
c_{i\sigma }= \sum_{\vec{k}}
\frac{1}{\sqrt{N}}b_{\vec{k}\sigma}\exp(i\vec{k} \cdot \vec{r}_i),
\qquad i\in B
\end{eqnarray}

The summation over $\vec{k}=(k_x,k_y)$ runs in the reduced Brillouin
zone. $N$ is the number of sites in each sublattice. By using a
spinor representation, the effective Hamiltonian can be written as
\begin{eqnarray}
H_{eff}=\sum_{\vec{k}}\eta^\dagger_{\vec{k}}M_{\vec{k}}
\eta_{\vec{k}} + const.,
\end{eqnarray}
where
$\eta^{\dagger}_{\vec{k}}=(a^\dagger_{\vec{k}\uparrow},b^\dagger_{\vec{k}\uparrow},a_{-\vec{k}\downarrow},b_{-\vec{k}\downarrow})$
and the matrix $M_{\vec{k}}$ reads
\begin{eqnarray}\label{matrix}
 M_{\vec{k}}=\left(%
\begin{array}{cccc}
A(\vec{k})-\mu & B(\vec{k})\\
B^{\dagger}(\vec{k})  & -A(\vec{k})+\mu  \\

\end{array}%
\right)
\end{eqnarray}
where both $A(\vec{k})$ and $B(\vec{k})$ are $2\times2$ matrices
whose elements are given by
\begin{eqnarray}
A_{11}&=&-(2g_t t'+ \frac{3}{4}g_s J' \xi_{a})cos{k_a} \nonumber \\
A_{22}&=&-(2g_t t'+ \frac{3}{4}g_s J' \xi_{b})cos{k_b} \nonumber \\
A_{12}&=&A_{21}= - 2 g_t t (cos{k_x}+cos{k_y})    \nonumber \\
              & & -  \frac{3}{4}g_s J (\xi_{x}cos{k_x} + \xi_{y}cos{k_y}) \nonumber \\
B_{11}&=&\frac{3}{4}g_s J' \Delta_{a} \cos{k_a} \nonumber\\
B_{22}&=&\frac{3}{4}g_s J' \Delta_{b} \cos{k_b}\nonumber \\
B_{12}&=&B_{21}=\frac{3}{4}g_s J(\Delta_{x} \cos{k_x} + \Delta_{y}
\cos{k_y})\nonumber
\end{eqnarray}
with $k_{\tau}=\vec{k}\cdot \vec{\tau}$. The matrix $M_{\vec{k}}$
can be diagonalized by employing an unitary transformation as
\begin{eqnarray}\label{cgam}
 &\,&\left[
{\begin{array}{*{20}c}
   {\gamma _{\vec{k} \uparrow }^{1\dag } }, & {\gamma _{\vec{k} \uparrow }^{2\dag } }, & {\gamma _{-\vec{k} \downarrow }^{1} }, & {\gamma _{-\vec{k} \downarrow }^{2 }}  \\
\end{array}} \right] =\nonumber\\ &\,&\left[ {\begin{array}{*{20}c}
   {a^\dagger_{\vec{k}\uparrow}  }, & {b^\dagger_{\vec{k}\uparrow} }, & {a_{-\vec{k}\downarrow} }, & {b_{-\vec{k}\downarrow} }  \\
\end{array}} \right]\left[ {\begin{array}{*{20}c}
   {u_1^1 } & {u_1^2 } & { - v_2^{1 * } } & { - v_1^{2 * } }  \\
   {u_2^1 } & {u_2^2 } & { - v_2^{1 * } } & { - v_2^{2 * } }  \\
   {v_1^1 } & {v_1^2 } & {u_1^{1 * } } & {u_1^{2 * } }  \\
   {v_2^1 } & {v_2^2 } & {u_2^{1 * } } & {u_2^{2 * } }  \\
\end{array}} \right] \nonumber
\\ \nonumber
\end{eqnarray}
where the $\vec{k}$ dependence of $u^n_i$ and $v^n_i$ are implied.
The matrix element  $u$ and $v$ satisfy the following conditions:
\begin{eqnarray}\label{m3}
\begin{array}{l}
 \sum\limits_i {u_i^1 } u_i^{2*}  + v_i^1 v_i^2  = 0 \qquad  \sum\limits_i { - u_i^1 } v_i^2  + v_i^1 u_i^2  = 0 \\
 \sum\limits_i {u_1^n } u_2^{n*}  + v_1^{n*} v_2^n  = 0 \quad \sum\limits_i {u_1^n } v_2^{n*}  - u_1^n v_2^n  = 0
 \end{array} \nonumber
\end{eqnarray}
In terms of $u$'s and $v$'s the self-consistent equations for eight
mean field and the hole density are given by,
\begin{eqnarray}
\xi _{x(y)}  &=& {2 \over \textstyle N}\sum\limits_{k,n} {[fu_1^{n*}
u_2^n  + (1 - f)v_1^n v_2^{n*} ]\cos k_{x(y)}}, \nonumber \\
\xi _a &=&{2 \over \textstyle N}\sum\limits_{k,n} {[f|u_1^n |^2  +
(1 - f)|v_1^n |^2 ]\cos k_a }, \nonumber \\ \xi _b &=& {2 \over
\textstyle N}\sum\limits_{k,n}
                     {[f|u_2^n |^2  + (1 - f)|v_2^n |^2 ]\cos
                     (k_b)},  \nonumber \\
\Delta _{x(y) }  &=&{2 \over \textstyle N}\sum\limits_{k,n} {[(1 -
f)u_1^n v_2^{n*}  - fv_1^{n*} u_2^n ]\cos k_{x(y)}},  \nonumber \\
\Delta _a &=& {2 \over \textstyle N}\sum\limits_{k,n} {(1 - 2f)u_1^n
v_1^{n*} \cos {k_a} }, \nonumber \\  \Delta _b       &=& {2 \over
\textstyle N}\sum\limits_{k,n} {(1 - 2f)u_2^n v_2^{n*} \cos k_b},
\nonumber \\  1-\delta        &=&{2 \over \textstyle
N}\sum\limits_{k,n}{f|u_2^n |^2  + (1 - f)|v_2^n |^2 }, \nonumber
\end{eqnarray}
where the band index $n$ runs over $1,2$ and $f$ is the Fermi
distribution function $f(E_n)$ which is step function at zero
temperature. The diagonalization of $M_{\vec{k}}$ and the
self-consistent equations are carried out numerically to obtain the
phase diagram and the results reported in the text.


\begin{references}

\bibitem{book} H.T. Diep, Frustrated Spin Systems (World Scientific Publishing Co., Singapore, 2004).

\bibitem{Aoki} H. Aoki, J. Phys. Cond. Matt. \textbf{16}, V1 (2004).

\bibitem{Anderson87} P.W. Anderson, Science \textbf{235}, 1196 (1987).

\bibitem{VBC} N. Read and S. Sachdev, Phys. Rev. Lett. {\bf 62} 1694 (1989).

\bibitem{NaCoO} K. Takada, H. Sakurai, E. Takayama-Muromachi, F. Izumi,
R.A. Dilanian and T. Sasaki, Nature \textbf{422}, 53 (2003).


\bibitem{Org1} Y. Shimizu, K. Miyagawa, K. Kanoda, M. Maesato
and G. Saito, Phys. Rev. Lett. {\bf 91}, 107001 (2003).

\bibitem{Kagome} P. Mendels, A. Keren, L. Limot, M. Mekata, G. Collin and M.
Horvatic, Phys. Rev. Lett. {\bf 85}, 3496 (2000).


\bibitem{pyro1} S. Yonezawa, Y. Muraoka, Y. Matsushita and Z. Hiroi, J. Phys.
Cond. Matt. \textbf{16}, L9 (2004).

\bibitem{pyro2} Y. Kasahara, Y. Shimono, T. Shibauchi, Y. Matsuda, S. Yonezawa, Y. Muraoka, and Z.
Hiroi, Phys. Rev. Lett. \textbf{96}, 247004 (2006).


\bibitem{Fuji} S. Fujimoto,  Phys. Rev. Lett. \textbf{89}, 226402 (2002); Phys. Rev. B {\bf 67}, 235102 (2003).

\bibitem{paradigm} R.R.P. Singh, O.A. Starykh, and P.J. Freitas,
J.\ Appl.\ Phys.\ {\bf 83}, 7387 (1998).

\bibitem{olegt} O. Tchernyshyov, O.A. Starykh, R. Moessner, and A.G. Abanov,
Phys. Rev. B  {\bf 68}, 144422 (2003).


\bibitem{hermele} M. Hermele, M.P.A. Fisher, and L. Balents, Phys. Rev.  B {\bf 69}, 064404 (2004).


\bibitem{Balents05}  O.A. Starykh, A. Furusaki, and Leon Balents, Phys. Rev. B {\bf 72}, 094416
(2005).

\bibitem{Ogata03} M. Ogata, J. Phys. Soc. Jpn. \textbf{72}, 1839 (2003).

\bibitem{d+id} B. Kumar and B.S. Shastry, Phys. Rev. B {\bf 68}, 104508 (2003);
G. Baskaran, Phys. Rev. Lett. {\bf 91}, 097003 (2003).

\bibitem{Kim04} C.H. Chung and Y.B. Kim, Phys. Rev. Lett. \textbf{93}, 207004
(2004).

\bibitem{palee} S.S Lee and P.A. Lee, Phys. Rev. Lett. \textbf{95}, 036403
(2005).

\bibitem{Gan06} J.Y. Gan, Y. Chen and F.C. Zhang, Phys. Rev. B \textbf{74}, 094515 (2006).

\bibitem{Poil1} A. Lauchli, D. Poilblanc, Phys. Rev. Lett. \textbf{92}, 236404 (2004).

\bibitem{Poil2} D. Poilblanc, Phys. Rev. Lett. \textbf{93}, 197204 (2004).

\bibitem{Poil3} D. Poilblanc, A. Lauchli, M. Mambrini, and F. Mila,
Phys. Rev. B {\bf 73}, 100403(R) (2006).


\bibitem{Zhang88}  F.C. Zhang, C. Gros, T.M. Rice and H. Shiba,  Supercond. Sci.
Tech. \textbf{1}, 36 (1988).


\bibitem{vannila} P.W. Anderson, P.A Lee, M. Randeria, T.M. Rice, N. Trivedi and F.C. Zhang,
J. Phys. Cond. Matt. {\bf 24}, R755 (2004).



\bibitem{gutzwiller} M.C. Gutzwiller, Phys. Rev. {\bf 137}, A1726(1965).

\bibitem{vallhardt} D. Vollhardt, Rev. Mod. Phys.{\bf 56}, 99 (1984).

\bibitem{kotliar} G. Kotliar and J. Liu, Phys. Rev. B {\bf 38}, 5142 (1988).


\end{references}
\end{document}